\documentclass[aps,prl,twocolumn,showpacs,groupedaddress]{revtex4}  
\usepackage{graphicx}  
\usepackage{dcolumn}   
\usepackage{bm}        
\usepackage{amssymb}   

\newcommand{\be}{\begin{equation}}
\newcommand{\ee}{\end{equation}}
\newcommand{\ba}{\begin{eqnarray}}
\newcommand{\ea}{\end{eqnarray}}
\newcommand{\ban}{\begin{eqnarray*}}
\newcommand{\ean}{\end{eqnarray*}}
\newcommand{\p}{\paragraph{}}

\newcommand{\ket}[1]{\mbox{$ | #1 \rangle $}}

\def\opone{\leavevmode\hbox{\small1\kern-3.8pt\normalsize1}}

\newcommand{\one}{\leavevmode\hbox{\small1\normalsize\kern-.33em1}}

\renewcommand{\quote}[1]{`#1'}

\newcommand{\fig}[1]{(Fig.~\ref{#1})}

\begin{document}

\title{Quantum Teleportation with a 3-Bell-state Analyzer}

\author{J.A.W. van Houwelingen\footnote[2]{email: jeroen.vanhouwelingen@physics.unige.ch}, N. Brunner, A. Beveratos, H. Zbinden, and N.
Gisin} \affiliation{Group of Applied Physics, University of
Geneva, Switzerland\\}
\date{\today}

\begin{abstract}
We present a novel Bell-state analyzer for time-bin qubits
allowing the detection of three out of four Bell-states with
linear optics, only two detectors and no auxiliary photons. The
theoretical success rate of this scheme is $50\%$. A teleportation
experiment was performed to demonstrate its functionality. We also
present a teleportation experiment with a fidelity larger than the
cloning limit.
\end{abstract}

 \maketitle \p A Bell-State
Analyzer (BSA) is an essential part of quantum communications
protocols such as a quantum relay based on quantum teleportation
 \cite{relay,relay1,relay2,TeleportationBennet}, entanglement
swapping \cite{SwappingGeneve} or quantum dense coding
 \cite{DensecodingTheo,DensecodingExp}. It has been shown that, using only linear optics, a
BSA for qubits has a maximal success rate of $50\%$ when no
auxiliary photons are used \cite{BSMlimit,KLM}. This, however,
does not limit the number of Bell-states one can measure, but only
the overall success rate. A complete BSA could be achieved using
either nonlinear optics or using continuous variable encoding
\cite{CWTeleportFullBellTheo,CWTeleportFullBellExp}. However, each
of these two alternatives carry some significant drawback. The
nonlinear optics approach has exceedingly low efficiency
\cite{NonLinearBell}; while continuous variable encoding has the
disadvantage that postselection is not possible. Note that
postselection is a very useful technique that allows one to use
only \quote{good} measurement results and straightaway eliminate
all others \cite{CVpostselection}.

Today's optimal BSA schemes based on linear optics for qubits are
only able to detect two out of four Bell-states
\cite{TeleportGeneve,TeleportVienne,SwappingGeneve,SwappingVienne}.
Here we present a novel scheme for a BSA which achieves the $50\%$
upper bound of success rate, but can distinguish three out of the
four Bell-states while using only two detectors. We demonstrate
this scheme in a quantum teleportation experiment at telecom
wavelengths. The new BSA is inspired by, although not limited to,
the time-bin implementation of qubits and is thus fully compatible
with the field of Quantum Communications \cite{OverviewQC}.

At first, one may think that detecting 3 out of 4 Bell-state
provides a full BSA. Indeed, if the BSA would consist of a
standard Von Neumann projective measurement, then the fourth
Bell-state would merely correspond to the non-detection of the 3
others. But our BSA is a new example of the power of generalized
quantum measurements, it uses a Positive Operator Valued
Measurement (POVM) with 21 possible outcomes. Some outcomes of
this POVM (see Fig. 1) correspond to one of the 3 Bell-states that
can be distinguished unambiguously and thus detect this state. The
others correspond to inconclusive results. More specifically, the
Bell-state $\ket{\psi_{+}}$ is always detected, $\ket{\phi_{-}}$
is never detected, while $\ket{\psi_{-}}$ and $\ket{\phi_{+}}$ are
detected with a 50\% success rate.

\begin{figure}
\includegraphics[scale=0.3]{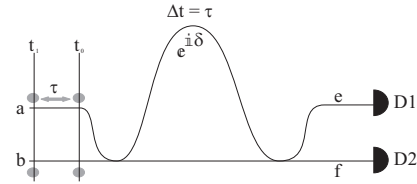}
\caption{\label{brunner}A schematic representation of the new type
of Bell-state Measurement. When two qubit states are sent into a
time-bin interferometer the output state is a mixture of photons
in two directional modes and three temporal modes. By looking at
certain combinations of these photons a Bell-state Measurement can
be performed for three different Bell-States}
\end{figure}

In previous BSAs the main method was to use a beamsplitter
followed by detectors to determine the input Bell-state. We
replace this simple approach by a time-bin interferometer
equivalent to the ones used to encode and decode time-bin qubits
\fig{brunner}
 \cite{OverviewQC}. The two photons of the Bell-state enter in port
$a$ and $b$, respectively. In a BSA using only a beamsplitter, one
is able to distinguish between $\psi_{\pm}$ but not $\phi_{\pm}$
since in the later case the two photons will experience photon
bunching but this interference does not contain the phase
information that distinguishes $\phi_{-}$ from $\phi_{+}$. In our
case, the first beamsplitter acts like above, but we introduce a
second interference possibility between the two photons on the
second beamsplitter. This second interference allows one to
distinguish more than just two Bell-states.

\begin{table*}
\begin{center}
\begin{tabular}{l|c c c c c c c c c c c c c c c c c c c c c c}
$D1$ & 00 &    & 11 &    & 22 &    & 01 &    & 02 &    & 12 &    & 0 & 2 & 1 & 1 & 0 & 1 & 2 & 0 & 2 \\
$D2$ &    & 00 &    & 11 &    & 22 &    & 01 &    & 02 &    & 12 & 0 & 2 & 1 & 0 & 1 & 2 & 1 & 2 & 0 \\
\hline \ket{\phi_{+}}
     &1/16&1/16&    &    &1/16&1/16&    &    &    &    &    &    &1/8&1/8&\textbf{1/2}&   &   &   &   &   &   \\
\ket{\phi_{-}}
     &1/16&1/16& 1/4& 1/4&1/16&1/16&    &    &    &    &    &    &1/8&1/8&   &   &   &   &   &   &   \\
\ket{\psi_{+}}
     &    &    &    &    &    &    & \textbf{1/8}& \textbf{1/8}&    &    & \textbf{1/8}& \textbf{1/8}&   &   &   &\textbf{1/8}&\textbf{1/8}&\textbf{1/8}&\textbf{1/8}&   &   \\
\ket{\psi_{-}}
     &    &    & 1/4& 1/4&    &    &    &    & \textbf{1/8}& \textbf{1/8}&    &    &   &   &   &   &   &   &   &\textbf{1/8}&\textbf{1/8}\\
\end{tabular}
\caption{\label{table:coincidances}The table shows the probability
to find any of the 21 possible coincidences as a function of the
input Bell-State. A \quote{0} in row D1 means that a photon was
found at detector \quote{D1} and at a time corresponding to the
photon having taken the short path in the interferometer and it
was originally a photon in time-bin $t_0$, a \quote{1} corresponds
to $t_0+\textbf{1}\times\tau$ with $\tau$ corresponding to a the
difference between the time-bins etc. Note that several
combinations of detection are possible for only one Bell-state(the
bold entries), therefor when such a combination is found a
Bell-state Measurement was performed. The theoretical probability
of a successful measurement is 0.5 which is the optimal value
using only linear optics \cite{BSMlimit}.}
\end{center}
\end{table*}

The output coincidences in ports $e$ and $f$, i.e. on detectors
$D1$ and $D2$, are summarized in table \ref{table:coincidances} as
a function of the input state. By convention, a coincidence at
time \quote{0} means that the photon did not accumulate any delay
with regards to its creation pulse. This is only possible if it
took the short path in the BSA and it was originally a photon in
time-bin $t_0$ (Fig. \ref{brunner}). A coincidence at time
\quote{1} means that the photon was originally in $t_1$ and took
the short path of the BSA interferometer or it was in $t_0$ and
took the long path. A coincidence at time \quote{2} then means the
photon was in $t_1$ and took the long path.

In table \ref{table:coincidances} one can distinguish two cases.
Either the result is unambiguous in which case we have
successfully distinguished a Bell-state. Or the result could have
been caused by two specific Bell-states, i.e. the result is
ambiguous and hence inconclusive.

The above described approach is correct in the case were the phase
$\delta$ of the BSA interferometer is set to 0. Experimentally, it
is not trivial to align this phase because the pump-interferometer
and the BSA interferometer involve different wavelengths. For an
arbitrary phase $\delta$ the situation is slightly different and
interesting. In such a case, our BSA still distinguishes 3
Bell-states, but these are no longer the Bell-states of the
computational basis. For a teleportation experiment this means the
basis for the measured Bell-states is not the same as the basis
for the entangled states shared between Bob and Charlie. Still,
perfect teleportation is possible, but with the difference that
the unitary transformations that Bob has to apply after receiving
the classical information about the result of the BSA have changed
and no longer include the identity: all unitary transformations
are non-trivial but they remain experimentally feasible. More
specifically, the analyzed Bell-states are:
\begin{eqnarray}
\phi^\prime_{\pm} &=& \ket{00} \pm e^{2i\delta} \ket{11} \label{eq:phipm}\\
\psi^\prime_{\pm} &=& e^{i\delta} (\ket{01} \pm \ket{10})
\label{eq:psief}
\end{eqnarray}
These new Bell-states are equivalent to the standard states except
that the $\ket{1}$ is replaced by $e^{i\delta}\ket{1}$ for each of
the input modes. Therefore the unitary transformations that have
to be applied to retrieve the original state of the teleported
photon also have to be modified from [$I$, $\sigma_z$, $\sigma_x$,
$\sigma_z \sigma_x$] to [$\sigma_{2\delta}$,
$\sigma_z\sigma_{2\delta}$, $\sigma_x$, $\sigma_z\sigma_x$].
Here $\sigma_{2\delta}=e^{-2i\delta}P_{\ket{1}}+P_{\ket{0}}$ 
is a phase shift of 2$\delta$ to be applied to the time bin
$\ket{1}$.

\begin{figure*}
\includegraphics[scale=0.8]{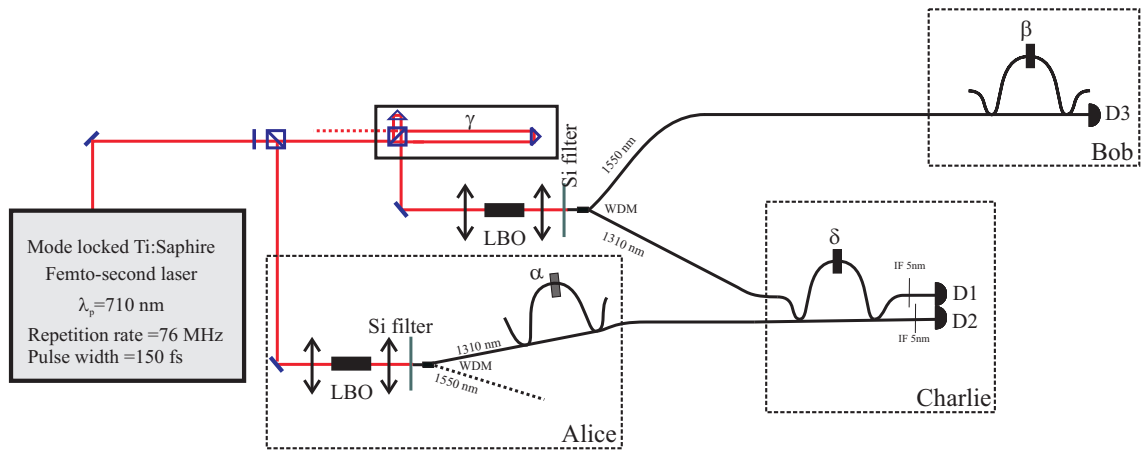}
\caption{\label{teleportbrunnerSETUP}A rough overview of the
experimental setup. The fiber interferometers shown here are in
reality michelson-interferometers, for the interferometer in the
BSA two circulators are used to have two separate inputs and
outputs. Not shown in the figure is the method used for
stabilizing the interferometers.}
\end{figure*}

In a more realistic experimental environment the success
probabilities of the BSA will be affected by detector limitations,
because existing photon detectors are not fast enough to
distinguish photons which follow each other closely (in our case
two photons separated by $\tau=1.2ns$) in a single measurement
cycle. This limitation rises from the dead time of the
photodetectors. When including this limitation we find that the
maximal attainable probabilities of success in our experimental
setup are reduced to $1/2$, $1/4$ and $1/2$ for $\psi_+$, $\psi_-$
and $\phi_+$ respectively. This leads to an overall probability of
success of $5/16$ which is greater by $25\%$ than the success rate
of $1/4$ with a BSA consisting only of one beamsplitter and two
identical detectors.

In order to demonstrate successful Bell-state analysis we
performed a teleportation experiment. A schematic of the
experimental setup is shown in Fig. \ref{teleportbrunnerSETUP}.
Alice prepares a photon in the state
$\ket{\zeta_A}=\ket{0}+e^{i\alpha}\ket{1}$. Bob analyzes the
teleported photon and measures interference fringes for each
successful BSA announced by Charlie. The setup consist of a
mode-locked Ti-sapphire laser creating 150 fs pulses with a
spectral width of 4nm, a central wavelength of 711nm and a mean
power of 400mW. This beam is split in two beams using a variable
coupler ($\lambda/2$ and a PBS). The reflected light (Alice) is
sent to a Lithium tri-Borate crystal(LBO, Crystal Laser) were by
parametric down-conversion a pair of photons is created at 1.31
and 1.55 $\mu$m. Pump light is suppressed with a Si filter, and
the created photons are collected by a single mode optical fiber
and separated with a wavelength-division-multiplexer (WDM). The
1.55 $\mu$m photon is ignored whereas the 1.31 $\mu$m is send to a
fiber interferometer which encodes the qubit on the equator of the
Bloch sphere. In the same way, the transmitted beam (Bob) is send
onto another LBO crystal after having passed through an unbalanced
Michelson bulk optics interferometer. The non-degenerate entangled
photons produced in this way corresponds to the $\phi_{+}$ state.
The photons at 1.31 $\mu$m are send to Charlie in order to perform
the Bell-state measurement. In order to assure temporal
indistinguishability, Charlie filters the received photons down to
a spectral width of 5 nm. In this way the coherence time of the
generated photons is greater than that of the photons in the pump
beam, and as such we can consider the photons to be emitted at the
same time. Bob filters his 1.55 $\mu$m photon to 15 nm in order to
avoid multi-photon events \cite{FourPhoton, FourPhoton2}. A liquid
Nitrogen cooled Ge Avalanche-Photon-Detector (APD) D1 with passive
quenching detects one of the two photons in the BSA and triggers
the InGaAs APDs (id Quantique) D2 and D3. Events are analyzed with
a time to digital converter (TDC, Acam) and coincidences are
recorded on a computer.

Each interferometer is stabilized in temperature and an active
feedback system adjusts the phase every $100$ seconds using
separate reference lasers. In this way the quantum teleportation
scheme works with independent units and is ready for \quote{in the
field} experiments. A more detailed description of the active
stabilization is given in ref \cite{SwappingGeneve}.

\begin{figure}
\begin{tabular}{l r}
\includegraphics[width=4cm]{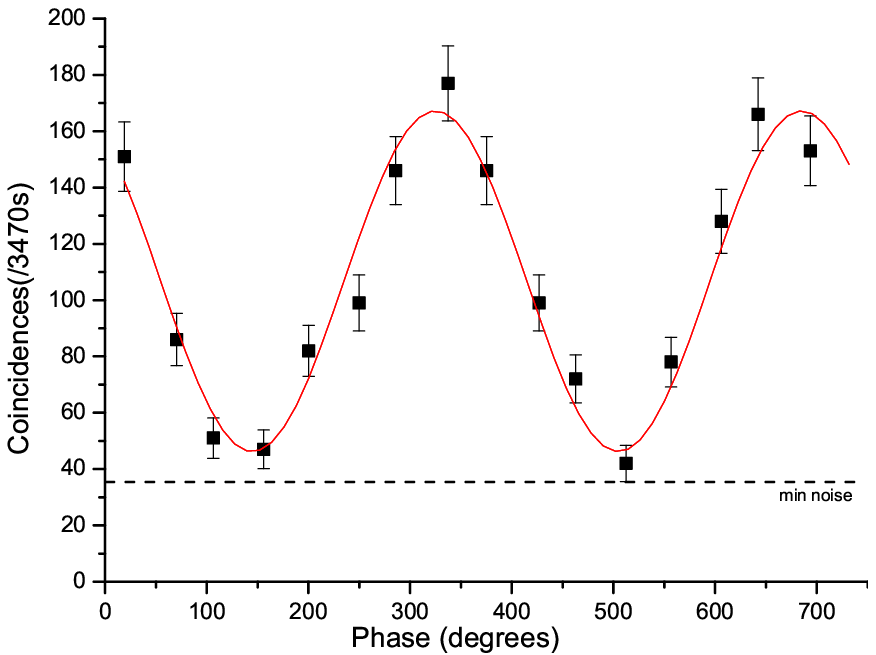}
&
\includegraphics[width=4cm]{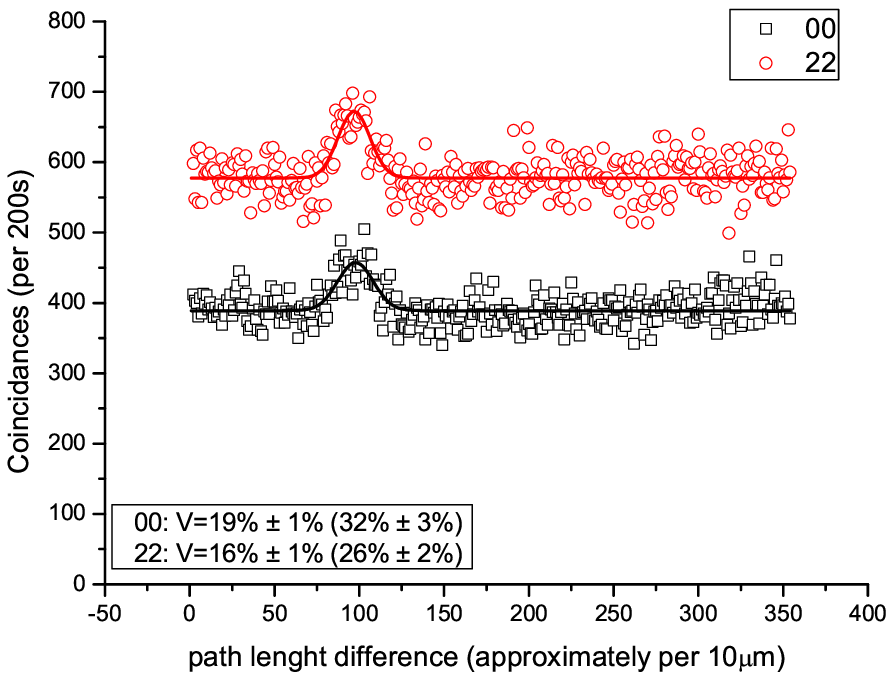}
\end{tabular}
\caption{\label{dip}\emph{Left:} The Result of the 1-Bell state
teleportation experiment (a beamsplitter instead of the
interferometer) with $F_{raw}=79\%$ and $F_{net}=91\%$.
\emph{Right:} A typical result of the scan in delay of the
coincidences SS and LL. The predicted antidip is clearly visible
in both curves with a visibility after noise substraction of
$32\%$ for \quote{00} and $26\%$ for \quote{22} (max=1/3). }
\end{figure}

The temporal indistinguishability of the photons arriving at the
BSA is usually tested by measuring a Hong-Ou-Mandel dip. In our
BSA this is not directly possible. Photons that have bunched on
the first interferometer will be split up by the second
interferometer in a nondeterministic manner and as such there will
be no decrease in the number of coincidences when looking at the
same time of arrival. There will actually be an increase for these
coincidences because the amount of photons that took a different
path in the interferometer will decrease. In our experimental
setup this means one has to look at an increase in coincidence
rate at times \quote{00} or \quote{22}. A typical result from this
alignment procedure can be found in \fig{dip}. The expected
antidips have a net visibility of $32\%$ and $26\%$ when
subtracting noise. The maximal attainable value is $1/3$ due to
undesired but unavoidable double-pair events
\cite{MandelDipGeneve}.

First we performed a quantum teleportation with independent units
using a beam splitter for the Bell-state measurement. This enables
us to test our setup in terms of fidelity. In order to check the
fidelity of the teleported sate, Bob sends the teleported photon
through an analyzing interferometer and measures the interference
fringes conditioned on a successful BSA. When Bob scans the phase
of his interferometer we obtained a raw visibility of $V=57\%$
($F=79\%$) and a net visibility of $V=83\% \pm 4$ ($F=91\% \pm 2$)
clearly higher than the cloning limit of F$=5/6$ \fig{dip}. We
then switched to the new BSA. This new setup introduces about 3dB
of excess loss, due to added optical elements including the
interferometer and its stabilization optics. These losses result
in a lower count rate. For experimental reasons we now scan the
interferometer of Alice instead of Bob. The experiments were
performed for approximately 4.4 hours per point in order to
accumulate enough data to have low statistical noise. The expected
interference fringes after Bob's interferometer are of the form
$1\pm \cos (\alpha + \beta - \gamma)$ for a projection on
$\psi_{\pm}$ and $1+\cos (\alpha - \beta + \gamma - 2*\delta)$ for
a projection on $\phi_{+}$. Hence one would expect to find three
distinct curves, two with a phase difference of $\pi$, and the
third dephased by $-2*(\beta - \gamma + \delta)$. Where $\alpha$,
$\beta$, $\gamma$ and $\delta$ are the phases of Alice, Bob, the
entangled pair preparation interferometer and the BSA measurement
interferometer respectively. Note that one can set the phase
$\gamma$ to $0$ as reference phase and that Bob is able to derive
the phase value $\delta$ of the BSA interferometer just by looking
at the phase differences between the fringes made by $\psi_{\pm}$
and $\phi_{+}$ and his knowledge about $\beta$.

\begin{figure}
\includegraphics[width=7.5cm]{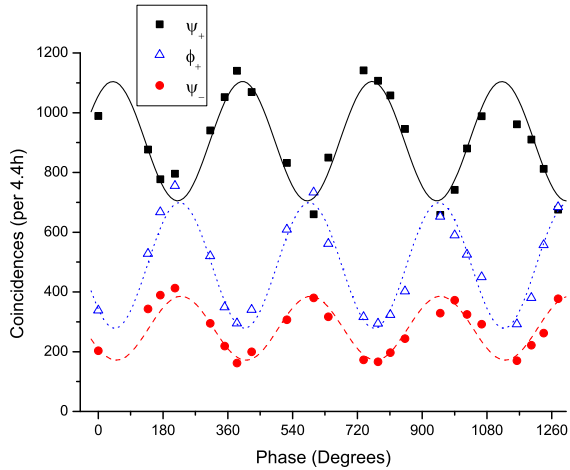}
\caption{\label{results} These graphs show the interference
fringes found when scanning the interferometer at Bob. we found
visibilities of 0.22,0.43 and 0.38 for $\ket{\psi_+}$,
$\ket{\phi_+}$ and $\ket{\psi_-}$. The average visibility of the
BSA is $V_{avg}=0.34$ (F=0.67). If we subtract the noise (not
shown) we find visibilities of 0.32,0.50 and 0.71 for
$\ket{\psi_+}$, $\ket{\phi_+}$ and $\ket{\psi_-}$ which leads to
an average of $V_{avg}=0.51$ ($F=0.76$) }
\end{figure}

\begin{figure}[t]
\includegraphics[width=7.5cm]{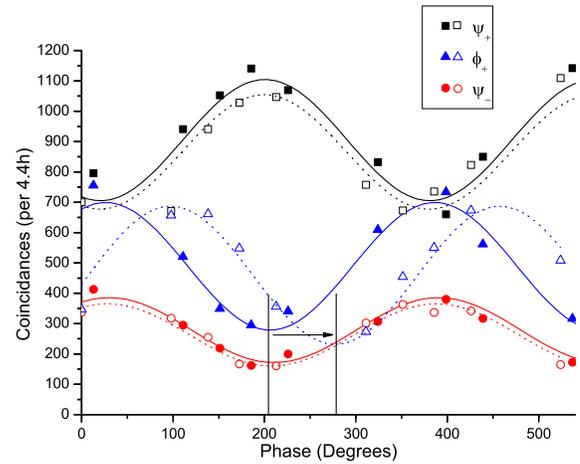}
\caption{\label{resultsdelta}This graph shows two measurements
that were taken with a different value of $\delta$. There is a
clear shift of the fringe for $\ket{\phi_+}$ with regards to the
other fringes. }
\end{figure}

In Fig. \ref{results} we show the coincidence interference fringes
between Bob and a successful BSA. As expected fringes for
$\ket{\psi_-}$ and $\ket{\psi_+}$ have a $\pi$ phase difference
due to the phase flip caused by the teleportation. On the other
hand fringes for $\ket{\psi_+}$ and $\ket{\phi_+}$ are dephased by
$-2(\beta+\delta)$ which in this case we had arranged to be
approximately $0$. The raw visibilities obtained for the
projection on each Bell-state are $V_{\psi_-}=0.38$,
$V_{\psi_+}=0.22$, $V_{\phi_+}=0.43$ which leads to an overall
value of $V=0.34$ ($F=0.67$). In order to check the dependence of
$\ket{\phi_+}$ on $\delta$ we also performed a teleportation with
a different value and we clearly observe the expected shift in the
fringe \fig{resultsdelta} while measuring similar visibilities.

Finally the authors would like to stress two points about this
novel BSA. First it is possible to implement this BSA for
polarization encoded photons by creating a second interference
possibility for H- and V-polarization. This would require 4
detectors, but an overall efficiency of $50\%$ can be achieved
with current day detectors. The limit of $50\%$ can also be
achieved with time-bin encoded qubits but would require the use of
ultra fast optical switches and two more detectors. We did not
implement this due to the losses associated with introducing
current day high speed integrated modulators. Secondly, even
though three out of four Bell-states can be distinguished, one can
not use this scheme in order to increase the limit of log$_23$
bits per symbol for quantum dense coding.

\p In conclusion we have shown experimentally that it is possible
to perform a three-state Bell analysis while using only linear
optics and without any actively controlled local operations on a
single qubit. In principle this measurement can obtain a success
rate of 50\%. We have used this BSA to perform a teleportation
experiment, and obtained a non-corrected overall fidelity of
$67\%$, after noise substraction we find F=$76\%$. Also, we
performed a teleportation experiment with a one state BSA which
exceeded the cloning limit.

\p The authors acknowledge financial support from the European
IST-FET project "RamboQ" and from the Swiss NCCR project "Quantum
Photonics" and thank C. Barreiro and J.-D. Gautier for technical
support.


\begin{thebibliography}{99}
\bibitem{CVpostselection} For experiments on the nature of
physics, such as Bell-tests, postselection can be a drawback. For
applications in Quantum Communication, however, it is advantageous
since it selects usefull measurement results.
\end{thebibliography}
\end{document}